\documentclass[a4paper,twocolumn,aps,prl,nolongbibliography,superscriptaddress,showpacs,showkeys,amsmath,amssymb,floatfix,nofootinbib]{revtex4-1}
\usepackage{lmodern}

\usepackage[T1]{fontenc}
\usepackage[utf8]{inputenc}
\usepackage{color}
\usepackage[unicode=true,pdfusetitle,
 bookmarks=true,bookmarksnumbered=true,bookmarksopen=true,bookmarksopenlevel=1,
 breaklinks=false,pdfborder={0 0 0},backref=false,colorlinks=true]
 {hyperref}
\hypersetup{
 citecolor=blue,filecolor=blue,linkcolor=blue,urlcolor=blue}
\usepackage[normalem]{ulem}
\makeatletter

\pdfpageheight\paperheight
\pdfpagewidth\paperwidth

 \@ifundefined{textcolor}{}
 {%
   \definecolor{BLACK}{gray}{0}
   \definecolor{WHITE}{gray}{1}
   \definecolor{RED}{rgb}{1,0,0}
   \definecolor{green}{rgb}{0,0.7,0}
   \definecolor{BLUE}{rgb}{0,0,1}
   \definecolor{CYAN}{cmyk}{1,0,0,0}
   \definecolor{MAGENTA}{cmyk}{0,1,0,0}
   \definecolor{YELLOW}{cmyk}{0,0,1,0}
 }

\renewcommand{\[}{\begin{equation}}
\renewcommand{\]}{\end{equation}} 
\usepackage{array}
\makeatother

\begin{document}

\title{Anisotropic stress as signature of non-standard propagation of gravitational waves}

\author{Ippocratis D.~Saltas}

\affiliation{School of Physics \& Astronomy, University of Nottingham, Nottingham,
NG7 2RD, United Kingdom}

\author{Ignacy Sawicki}

\affiliation{D\'epartment de Physique Th\'eorique and Center for Astroparticle Physics,
Universit\'e de Gen\`eve, Quai E. Ansermet 24, CH-1211 Gen\`eve 4, Switzerland}

\affiliation{African Institute for Mathematical Sciences, 6-8 Melrose Road, Muizenberg,
Cape Town, South Africa}

\author{Luca Amendola}

\affiliation{Intitut f\"ur Theoretische Physik, Ruprecht-Karls-Universit\"at Heidelberg,
Philosophenweg 16, 69120 Heidelberg, Germany}

\author{Martin Kunz}

\affiliation{D\'epartment de Physique Th\'eorique and Center for Astroparticle Physics,
Universit\'e de Gen\`eve, Quai E. Ansermet 24, CH-1211 Gen\`eve 4, Switzerland}

\affiliation{African Institute for Mathematical Sciences, 6-8 Melrose Road, Muizenberg,
Cape Town, South Africa}

\begin{abstract}
We make precise the heretofore ambiguous statement that anisotropic
stress is a sign of a modification of gravity. We show that in cosmological solutions of very general classes 
of models extending gravity --- all scalar-tensor
theories (Horndeski), Einstein-Aether models and bimetric massive
gravity --- a direct correspondence exists between perfect fluids apparently carrying anisotropic stress and a modification in the propagation of gravitational waves. Since the anisotropic stress can be measured in a model-independent
manner, a comparison of the behavior of gravitational waves from cosmological sources 
with large-scale-structure formation could in principle lead to new constraints on the theory of gravity. 
\end{abstract}
\maketitle

Over the last decade, we have established beyond reasonable doubt that, in its recent past, the expansion of the universe has been accelerating. This has challenged our beliefs about the theory of gravity: the only possibility available in general relativity with non-exotic matter is a cosmological constant, which would suffer from severe fine-tuning issues. Alternatively, the mechanism could be dynamical, i.e.\ feature at least one new degree of freedom. These dynamics would modify the predictions of concordance cosmology and give us a means to carry out precision tests of gravity at extremely large scales.

Frequently in extended models of gravity 
perfect fluids apparently carry anisotropic stress: there is gravitational slip, i.e. the values of the two scalar gravitational potentials sourced by matter are not equal. This affects structure formation and weak lensing. Recently, it was shown
that the ratio of the two potentials is actually a model-independent observable
\cite{Amendola:2012ky, Motta:2013cwa}, which Euclid should be able
to measure to a precision of a few percent, depending on the precise assumptions
\cite{Amendola:2013qna}. This begs the question as to what detecting or not detecting 
anisotropic stress actually means.

In this Letter we show that the
propagation of \emph{tensor} modes (gravitational waves, GWs) is also
modified whenever the anisotropic stress is present at first order in
perturbations sourced by perfect-fluid matter. We demonstrate this
relationship in the context of three very large
classes of extensions of the gravitational sector: general scalar-tensor
theories (Horndeski \cite{Horndeski:1974wa,Deffayet:2011gz}), Einstein-Aether
models \cite{Nordtvedt:1972zz,Will:1972zz,Hellings:1973zz} and bimetric massive gravity
\cite{Hassan:2011tf,Hassan:2011vm}. GWs are the only propagating degrees
of freedom in General Relativity, and it is natural to define modified gravity models
as those where the gravitational waves are modified in such a non--trivial manner. Since imperfect fluids with anisotropic stress also split the two gravitational potentials but do not modify the propagation of tensor modes, this definition allows us to separate modifications of gravity from imperfect fluids.

The emphasis of this paper is not on new calculations (see e.g.\ the review~\cite{Tsujikawa:2014mba}), but rather on new relations which are very general, were not noted before in the literature and could have a significant impact on tests of gravity on cosmological scales.

\section{Assumptions}

We assume that the universe is well-described by small linear perturbations
living on top of a spatially flat Friedmann metric. We take the line
element for the metric on which matter and light propagate as 
\[
\mathrm{d}s^{2}=a^2(\tau)\left( -(1+2\Psi)\mathrm{d}\tau^{2}+(1-2\Phi)\left[\delta_{ij}+h_{ij}\right]\mathrm{d}x^{i}\mathrm{d}x^{j}\right) \nonumber
\]
where $\tau$ is the conformal time, $a$ the scale factor, $\Phi$ and $\Psi$ are the scalar gravitational potentials
and $h_{ij}$ is the traceless spatial metric (tensor) perturbation,
i.e.~the gravitational wave. We assume that
the matter sector can be described as a fluid arising from the averaging of the motion of particles. We comment on the effect of this fluid's being imperfect. We use the prime to denote a derivative with respect to conformal time.

The presence of anisotropic stress results in a difference in values between the two scalar
potentials and can be described through the gravitational slip, 
\[
\eta\equiv\frac{\Phi}{\Psi}\,.
\]
In concordance cosmology, $\eta=1$, with small corrections appearing
from neutrino free-streaming. At second order in perturbations, anisotropic
stress also always appears even when the matter consists completely
of dust \cite{Ballesteros:2011cm}, but in the late universe should
be smaller than $\left|\eta-1\right|\lesssim10^{-3}$ \cite{Ballesteros:2011cm,Adamek:2013wja}.

On the other hand, various modifications of gravity (such as $f(R)$ \cite{Zhang:2005vt},
$f(G)$ \cite{Saltas:2010tt} or DGP \cite{Koyama:2005kd}) do feature an $\mathcal{O}(1)$ correction
to the slip parameter at linear order in perturbations, at least at
some scales and even in the presence of just a perfect-fluid matter. It is, however, well known that the value of $\eta$
can be modified by a change of frame, e.g.~a conformal rescaling
of the metric, making its value seemingly ambiguous.

In Refs~\cite{Amendola:2012ky,Motta:2013cwa}, it was shown that
comparing the evolution of redshift-space distortions of the galaxy
power spectrum with weak-lensing tomography allows us to reconstruct
$\eta$ as a function of time and scale in a model-independent manner.
Such an operational definition removes the frame ambiguity, since
the measurement picks out the particular metric on the geodesics of
which the galaxies and light move. It is the gravitational slip in
that metric that is being measured by such cosmological probes. With
the ambiguity of frame removed, the gravitational slip is a \emph{bona
fide }observable, rather than just a phenomenological parameter. Fixing
the metric also determines what is considered a gravitational wave:
we call these the propagating spin-2 perturbations of the metric on
which matter moves.\footnote{In the case of massive gravity, we are referring to 
the helicity-2 modes of the metric coupled to matter.}

In this Letter, we assume that the gravitational sector is extended
by one of three classes of models featuring a single extra degree
of freedom: (1) a very general scalar-tensor theory belonging to the
Horndeski class \cite{Horndeski:1974wa};
(2) Einstein-Aether theory featuring an extra vector and spontaneous
violation of Lorentz invariance; or (3) bimetric massive gravity. We will discuss each
of these in turn and show that similar conclusions hold.

\section{Modified Gravity Defined}\label{sec:MGdef}
Dynamical models of late-time acceleration can feature interactions between
the new degree of freedom and curvature/metric (scalar-tensor/Einstein-Aether) and the two
metrics (bimetric). On a cosmological background, these interactions can alter the 
speed of propagation of gravitational waves ($c_{\text{T}}$), make
the effective Planck mass ($M_{*}$) evolve in time  \cite{Riazuelo:2000fc} or add a mass $\mu$, giving
\begin{equation}
h''_{ij}+(2+\nu)H h'_{ij} + c_\text{T}^2 k^2h_{ij}+a^2\mu^{2}h_{ij}=a^2\Gamma\gamma_{ij}\,,\label{eq:GWeq}
\end{equation}
where $h_{ij}$ is the tensor wave amplitude in either of the two polarizations, $H\equiv a'/a$ is the Hubble rate in conformal time. The deviations away from standard behavior are contained in $\nu \equiv H^{-1}\frac{\mathrm{d}\ln M_{*}^{2}}{\mathrm{d}t}$, the \emph{Planck mass run rate}, 
and $c_\text{T}$, the speed of tensor waves, with both of these quantities defined in the Jordan frame of the matter.%
\footnote{Note that no observable quantity depends on $M_{*}$ itself, since a changed
Planck mass can always be reabsorbed into the definition of masses
if it is constant.} %
We will show that scalar-tensor and Einstein-Aether models can change $\nu$ and $c_\text{T}$. On the other hand, in massive bigravity, the equation is modified by the mass of the graviton $\mu$. The transverse-traceless tensor $\gamma_{ij}$ is a source term for the gravitational waves. In the case of bimetric massive gravity, $\gamma_{ij}$ is the gravitational wave in the second metric and the two tensor modes mix as they propagate. When the matter fluid has anisotropic stress, this appears as the source term $\gamma_{ij}$, but it never modifies the homogeneous part of Eq.~\eqref{eq:GWeq}. However, this anisotropic stress is itself coupled to the gravitational waves and can lead to dissipation for horizon-scale GW modes \cite{Durrer:1997ta,Weinberg:2003ur}.

As we stressed above, Eq.~\eqref{eq:GWeq} describes the evolution of the gravitational waves of the Jordan-frame metric. This choice is unique if our observations (e.g.\ redshifts, time delays) are taken to result from the geometry of the Universe. We should also note that, for bimetric massive gravity, the Einstein frame with standard gravitons does not exist even on a perturbative level. On the other hand, the issue of which of the two metrics matter couples to is an important one, which has to be fixed to define the model properly.  

As is frequently said, anisotropic stress is a feature of modified
gravity. For any gravity theory at the linear level, the anisotropy constraint
in the Newtonian gauge takes the form
\begin{align}
\Phi - \Psi = \sigma(t)  \Pi + \pi_\textrm{m}, \label{eq:AnisoEq-Gen}
\end{align}
with $\Pi$ a function of a  particular combination of background and linear perturbation variables, depending on the theory. The quantity $\sigma(t)$ is a background function only, depending on the parameters of the Lagrangian. The $\pi_\textrm{m}$ is the scalar anisotropic stress sourced by the matter fluid. This appears whenever the perfect-fluid approximation breaks down and the particle distribution contains higher moments than those described by a perfect fluid. For example, free-streaming in neutrinos gives such a term even in concordance cosmology, but such contributions are very small in the late universe.

The aim of this Letter is to provide  an unambiguous definition of modified gravity as one where the propagation
of \emph{gravitational waves} \eqref{eq:GWeq} is affected. The gravitational slip and gravitational waves are connected since both the anisotropy constraint \eqref{eq:AnisoEq-Gen} and the GW evolution equation \eqref{eq:GWeq} arise from the spatial--traceless part of the linearized Einstein equations. In the remainder of this Letter, we will demonstrate that the coupling $\sigma(t)$ appearing in the anisotropy equation (\ref{eq:AnisoEq-Gen}) consists of the quantities that also control the modification of the tensor propagator. This means that modified gravity models popular in the literature are included in our definition.

However, imperfect-fluid matter while acting as a source to both the anisotropy constraint \eqref{eq:AnisoEq-Gen} and the GW equation \eqref{eq:GWeq}, cannot directly modify the homogeneous part of the GW equation. Our definition of modified gravity therefore breaks the ambiguity that arises in the presence of such a source and points to an approach for differentiating modified gravity from imperfect fluids.

\section{Scalar-Tensor Theories}

In this section, we consider the most general class of theories featuring
one extra scalar degree of freedom which has Einstein equations with no more
than second derivatives on any background and are universally coupled
to matter: the Horndeski class of models.%
\footnote{We have not considered in detail the extension discussed in \cite{Gleyzes:2013ooa,Gleyzes:2014dya, Gao:2014soa}, where higher derivatives appear in the Einstein equations, but can
be eliminated by solving the constraints.}
This class includes the majority of the popular models of late-time
acceleration such as quintessence, perfect fluids, $f(R)$ gravity,
$f(G)$ gravity, kinetic gravity braiding and galileons (see e.g.\ the reviews \cite{Amendola2010,Clifton:2011jh}). The Horndeski Lagrangian is
defined as the sum of four terms that are fully specified by a non-canonical kinetic term $K(\phi,X)$
and three arbitrary coupling functions $G_{3,4,5}(\phi,X)$,
where $X=-g_{\mu\nu}\phi^{,\mu}\phi^{,\nu}/2$ is the canonical kinetic
energy term and where the comma denotes a partial derivative.

We make extensive use of the formulation for linear structure formation
in scalar-tensor theories introduced in Ref.~\cite{Bellini:2014fua}.
It was shown there that the form of linear perturbation equations for all Horndeski models
can be completely described in terms of the background expansion history,
density fraction of matter today $\Omega_{\text{m}0}$, and four independent
and arbitrary functions of time only, $\alpha_{\text{K}},\alpha_{\text{B}},\alpha_{\text{M}}$
and $\alpha_{\text{T}}$, which mix the four functional degrees of
freedom of the action, $K$ and $G_{i}$. The \emph{Planck mass run rate} $\alpha_{\text{M}}$
and the \emph{tensor speed excess} $\alpha_{\text{T}}$ control the existence of anisotropic stress. 
Unrelated to the anisotropic stress, if the \emph{braiding} $\alpha_{\text{B}}\neq 0$,
then the dark energy will cluster at small scales, with the \emph{kineticity} $\alpha_{\text{K}}$ controlling at what scales this happens. 

The anisotropy constraint in the notation of Eq.~\eqref{eq:AnisoEq-Gen} is \cite{DeFelice:2011hq}
\begin{align}
	\sigma &= \alpha_\text{M}-\alpha_\text{T}\, \label{eq:anisoeq}\\
	\Pi &=H\delta\phi/\dot{\phi} + \alpha_\text{T}/(\alpha_\text{M}-\alpha_\text{T})\Phi\,. \notag
\end{align}
where $\delta\phi$ is a perturbation of the scalar field. Note that the split between $\sigma$ and $\Pi$ above is arbitrary. The gravitational wave equation~\eqref{eq:GWeq} is modified through
\begin{align}
	&\nu = \alpha_\text{M}\,, &&c_\text{T}^2=1+\alpha_\text{T}\,, \\
	&\mu^2 = 0\,, &&\Gamma=0\,. \notag
\end{align}
It is clear from eq.~\eqref{eq:anisoeq} that when both $\alpha_{\text{M}}=\alpha_{\text{T}}=0$
there is no new contribution to either to anisotropic stress or tensor propagation. In the context of
scalar-tensor models and the late universe with $\pi_\textrm{m}\approx0$, a detection of anisotropic stress therefore
is direct evidence that one or both of the parameters $\alpha_{\text{T}}$
and $\alpha_{\text{M}}$ are different from their concordance values
of zero and that gravity is modified in the sense of this work. 

In principle, one could imagine that there may exist models defined
by a choice of the functions $\alpha_{i}$ in which the scalar perturbation arranges
itself dynamically in such a configuration that no gravitational slip
appears, even though one of $\alpha_\text{M,T}$ is not zero. 
This would be a very particular situation or one requiring a very tuned choice of model parameters. For example, it happens at the asymptotic future --- and static --- pure de-Sitter limit. It can be shown 
that it is in fact impossible to have such a cancellation in a model where the scalar has real dynamics. We defer the proof to a more technical follow-up study.

\section{Einstein-aether Theories}

Einstein-aether models \cite{Jacobson:2000xp, Jacobson:2008aj} are a class of theories which feature an extra vector degree of freedom (the \emph{aether}) $u^\mu$. They are a subclass of general vector theories requiring that $u^\mu$ be given a constant and timelike vacuum expectation value $u_{\mu} u^{\mu} = -1$ and that it be minimally coupled. This chooses a preferred frame, violating Lorentz symmetry. The infrared limit of Hořava-Lifshitz (HL) models \cite{Horava:2009uw, Blas:2009qj,Blas:2010hb} --- relevant for late-time cosmology --- is closely related, with the vector field forced to be hypersurface orthogonal and thus providing a natural slicing for the space-time \cite{Jacobson:2013xta}.

The Lagrangian can be written in a basis of four operators, through a kinematic decomposition of $\nabla_\mu u_\nu$ \cite{Jacobson:2013xta}: the squares of acceleration, expansion, twist and shear, and their associated dimensionless coefficients $c_{a}$, $c_{\theta}$, $c_{\omega}$ and $c_{\sigma}$, respectively.\footnote{In the language of Ref.~\cite{Jacobson:2013xta}, these correspond to $c_{a} \equiv -c_1 + c_4$, $c_{\theta} \equiv \frac{1}{3}(c_{1} + c_{3}) + c_{2}$, $c_\omega=c_1  - c_3$, $c_{\sigma} \equiv  c_{1} + c_{3}$.}

The extra dynamical degree of freedom at the linear level is the perturbation of the spatial components $u^{i}$ of the vector $u^{\mu}$, which can be decomposed into longitudinal and transverse parts as $u^{i} = \partial^{i} u + \hat{u}^{i}$. The longitudinal part modifies the anisotropy constraint \cite{Lim:2004js}, 
which in the notation of Eq.~\eqref{eq:AnisoEq-Gen} is
\begin{align}
 \Pi = \left( \frac{u}{a^2} \right)' \qquad \sigma = -c_{\sigma}\,.
\end{align}
At the same time, the parameters of the tensor equation \eqref{eq:GWeq} are given by
\begin{align}
&\nu = 0\,, &&c_\text{T}^2 = (1 + c_{\sigma})^{-1}\,, \\
&\mu^2 =0\,, &&\Gamma=0\,. \notag
\end{align}

In conclusion, the modifications of both the anisotropy constraint and the tensor wave equation are driven by the same coupling $c_\sigma$ of the shear. If $c_\sigma$ appears in the action, it will modify both the anisotropic stress and the gravitational wave propagation. Thus a detection of anisotropic stress in the late Universe with $\pi_\textrm{m}\approx0$ in the context of these models also implies that gravity is modified in the sense of this work.

\section{Bimetric Massive Gravity}

The bimetric massive gravity model features two dynamical metrics, $g_1$ and $g_2$, each with its own Einstein--Hilbert term in the action. In addition, a potential term describes non--derivative interactions between the two metrics, $U(g_{1}, g_{2}; a_{i})$. The five constants $a_i$ parametrize these interactions and are the theory's free parameters. The interactions inevitably give mass to one of the two metrics \cite{Boulanger:2000rq}, and the theory in general propagates a massless and a massive spin-two field \cite{Hassan:2011zd, Hassan:2011zd}, and it provides a non--linear extension of the Fierz-Pauli theory \cite{Fierz:1939ix}, which is free of the so-called Boulware-Deser ghost \cite{deRham:2010ik,deRham:2010kj,Hassan:2011hr,Hassan:2011tf,deRham:2011rn}. One usually considers the matter fields to be coupled to one of the metrics, which we shall call $g_{1}$.

Bimetric gravity provides a natural extension of the so-called dRGT massive gravity, with the latter being a subcase of the former, in the limit where the second metric becomes non-dynamical. Cosmological solutions for dRGT and bimetric theories have been studied in, for instance, \cite{Gumrukcuoglu:2011ew,Gumrukcuoglu:2011zh,Hassan:2011vm,Fasiello:2012rw,Langlois:2012hk,Comelli:2011zm} and \cite{vonStrauss:2011mq,2013JHEP...03..099A,Konnig:2013gxa,Konnig:2014dna,DeFelice:2013nba,DeFelice:2014nja,DeFelice:2014nja} respectively, with the aim of explaining the current acceleration of the universe without the need of an explicit cosmological constant in the action.
It has been shown however that in dRGT, homogeneous and isotropic backgrounds are not solutions of the background equations of motion \cite{D'Amico:2011jj}, or when these solutions exist, they suffer from strong coupling \cite{Gumrukcuoglu:2011zh}, ghost \cite{Fasiello:2013woa, DeFelice:2012mx} or non-linear instabilities \cite{DeFelice:2012mx,DeFelice:2013awa}, and we will therefore concentrate on the bimetric version only.\footnote{In fact, a gradient instability for the new helicity-0 mode in the bimetric setup appears to exist for for some choices of parameters \cite{Comelli:2012db,DeFelice:2014nja} but not others \cite{Konnig:2014xva}. Whenever healthy solutions exist, the conclusions of this Letter hold.}

We use the setup and notation of Ref.~\cite{Comelli:2012db} \footnote{For a similar analysis see also Ref. \cite{Solomon:2014dua}.}, choosing both the background metrics to be homogeneous and isotropic. At the linear level, the theory predicts the existence of anisotropic stress for the scalar Newtonian potentials of the matter metric $g_1$, giving the anisotropy constraint the form
\begin{align}
	\sigma =a^2 m^2 f_1\qquad \Pi = E_2
\end{align}
in the notation of Eq.~\eqref{eq:AnisoEq-Gen}. $E_2$ is the scalar coming from the tensor perturbation of the second metric $g_2$. The function $f_1$ is a background-dependent function that depends on the ratio between the scale factors of the two metrics and the constant parameters $a_i$. 

The equation for gravitational waves \eqref{eq:GWeq} is modified through
\begin{align}
&\nu = 0\,, &&c_\text{T}^2=1\,, \\
& \mu^2 = m^2 f_1\,, 	&&\Gamma = m^2 f_{1} \,. \notag
\end{align}
Massive bigravity models change neither the Planck mass nor the speed of gravitational waves. They do give gravitons a mass and an interaction term. As we can easily see, the coefficients modifying the anisotropy constraint and the graviton equation of motion are all proportional to $m^2 f_1$. Yet again, if anisotropic stress is observed in the late Universe with $\pi_\textrm{m}\approx0$ in the context of these models, we must conclude that gravity is modified in the sense of this work.

\section{Conclusions and Implications}

In this Letter, we have shown that a very close relationship exists
between two properties of general extensions of gravity which until 
now have not been considered together:
when anisotropic stress is apparently sourced by perfect-fluid matter perturbations at linear level, the propagation of gravitational waves is modified. Such a relationship generally exists
in all Horndeski theories with an extra scalar, Einstein-Aether theories featuring
an extra vector field and bimetric massive gravity, featuring a second
rank-2 tensor field --- this covers a very large fraction of all the extensions of gravity with homogeneous backgrounds. We conjecture that this is a feature of all 
models \emph{in general configurations} and we choose to use this physics as the unambiguous
definition of modified gravity.

We note here that the anisotropic stress and clustering of the new degree of freedom --- frequently described as a change to the effective Newton’s constant --- are both completely independent quantities, the presence of which is not contingent on each other.

The relationship between tensor propagation and gravitational slip is a result of both being part of the spatial-traceless part of the linearised Einstein equations: the same corrections in the action modify the anisotropy constraint and the action for the graviton. 

We stress that this relationship would hold whenever gravity is modified, not only at low redshifts where extensions
of gravity are frequently utilized as dynamical models of acceleration. For example, during recombination, if models of gravity with apparent anisotropic stress from perfect fluids are introduced, one would then need to adjust the behavior of gravitational waves. At the same time, this new anisotropic stress would change the lensing and the integrated Sachs-Wolfe effect. All these effects would modify the CMB spectrum, in particular the B-mode polarization \cite{Amendola:2014wma,Audren:2013dwa,Lim:2004js}.

This deep relationship between anisotropic stress and tensor modes implies that measurements
of large-scale structure and of gravitational waves can give independent information on the properties of each other.  For example, a comparison between the time of arrival of
neutrinos and gravitational waves from some energetic event 
is a probe of the speed of tensor modes $c_\text{T}$ and their mass $\mu$ \cite{Nishizawa:2014zna}. A luminosity distance from standard sirens imputed from the decay of the amplitude of the gravitational waves probes $\nu, \mu$ and $\Gamma$ \cite{Cutler:2009qv}.
Such observations are clearly extremely challenging and futuristic, but
may one day be possible.\footnote{Tests such as the binary pulsar \cite{Yagi:2013ava} probe the coupling of matter sources to gravitational waves and therefore are not necessarily sensitive to the modification in propagation described here.} On the other hand, the slip parameter $\eta$ in some models can be an order-one ratio of small numbers (e.g.~in $f(R)$ gravity, where the permitted parameter values are $\alpha_\text{M}=-\alpha_\text{B} \lesssim 10^{-5}$ \cite{Lombriser:2010mp}, while  $\eta=1/2$ inside the Compton scale). Measurements of anisotropic stress can be more informative about tensor modes than direct probes of gravitational waves in such a case. Ultimately, it should be possible to combine them to disambiguate the various properties of the theory of gravity at cosmological scales. We leave the discussion of how feasible this is to
future work.

\begin{acknowledgments}
\emph{Acknowledgements.}  We are grateful to Mariele Motta for contribution at initial stages of this work. Furthermore, we would particularly like to thank Emilio Bellini, Diego Blas, Marco Crisostomi and Emir G{\" u}mr{\" u}k{\c c}{\" u}o{\u g}lu for many valuable discussions and communication. We also thank the anonymous referees for suggestions that have improved this manuscript. The work of L.A.~is supported
by the DFG through TRR33 ``The Dark Universe''. M.K.~acknowledges
funding by the Swiss National Science Foundation. I.S.~is supported by the Marie Skłodowska-Curie Intra-European Fellowship Project ``DRKFRCS''. I.D.S. is supported by an
STFC consolidating grant. \end{acknowledgments}
\bibliographystyle{utcaps}
\bibliography{AnisoRefs}

\end{document}